\documentclass{iaus}
\usepackage{graphicx}

\title[Constraints on a strong X-ray flare in MCG-6-30-15] {Constraints on a
  strong X-ray flare in the Seyfert galaxy MCG-6-30-15}

\author[Goosmann \etal] {R.~W. Goosmann$^1$, B. Czerny$^2$, V. Karas$^1$,
  M. Dovciak$^1$,\\ G. Ponti$^3$, \and M. Mouchet$^4$}

\affiliation{
  $^1$~Astronomical Institute, Academy of Sciences, Bo{\v c}n{\' i}~II~1401,
       14131 Prague, Czech Rep.\\
  $^2$~Copernicus Astronomical Center, Bartycka 18, 00-716 Warsaw, Poland\\
  $^3$~Dipartimento di Astronomia, Universit\`a di Bologna, Via Ranzani 1,
       40127, Bologna, Italy\\\
  $^4$~Laboratoire ApC, Universit\'e Denis Diderot, 2 place Jussieu,
       75251 Paris Cedex 05, France\\}

\pubyear{2006}
\volume{238}
\pagerange{001--999}
\date{??? and in revised form ???}
\jname{Black Holes: from Stars to Galaxies -- across the Range of Masses}
\editors{V. Karas \& G. Matt, eds.}
\begin{document}

\maketitle

\begin{abstract}
We discuss implications of a strong flare event observed in the
Seyfert galaxy MCG-6-30-15 assuming that the emission is due to
localized magnetic reconnection. We conduct detailed radiative
transfer modeling of the reprocessed radiation for a primary source
that is elevated above the disk. The model includes relativistic
effects and Keplerian motion around the black hole. We show that for
such a model setup the observed time-modulation must be intrinsic to
the primary source. Using a simple analytical model we then
investigate time delays between hard and soft X-rays during the
flare. The model considers an intrinsic delay between primary and
reprocessed radiation, which measures the geometrical distance of the
flare source to the reprocessing sites. The observed time delays are
well reproduced if one assumes that the reprocessing happens in magnetically
confined, cold clouds.

\end{abstract}

The Seyfert galaxy MCG-6-30-15 was observed for 95 ksec with {\it
XMM-Newton} in the year 2000 (\cite[Wilms \etal\
2001]{wilms2001}). The X-ray lightcurve of this observation reveals a
strong flare event of which \cite[Ponti \etal\ (2004)]{ponti2004}
conducted a detailed analysis. In this proceedings note we discuss the
possibility that the flare is produced by localized magnetic
reconnection and we constrain some details of such a flare setup.

\section{Model lightcurves for an elevated flare source}  
\label{sec:LCs}

We imagine that the strong flare in MCG-6-30-15 originates in a
compact reconnection site elevated to a height $H$ above the surface of the
accretion disk. The radiation from this primary source partly shines
toward the disk and creates a hot spot. The distance of the spot's
center to the disk center is denoted by $r$. The flare is supposed to
be in Keplerian co-rotation with the disk and we assume that the primary
illumination sets on and fades out instantaneously. The irradiation of
the disk then evolves across the hot spot, starting from the spot
center and progressing toward the border. Therefore we expect the
lightcurve of the reprocessed radiation to be curved even if the time
evolution of the primary is box-shaped.

We want to model the exact shape of the lightcurve expected from such
a flare setup at different orbital phases of the disk. We first
conduct detailed local radiative transfer computations. The varying
intensity of the irradiation across the hot spot is taken into account
as well as the angular dependence of the reprocessed emission. The
vertical structure of the disk is assumed to remain in the same
hydrostatic equilibrium as before the onset of the flare. The profile
is computed with an extended version of the code described in
\cite[R{\'o}{\.z}a{\'n}ska \etal\ (2002)]{rozanska2002}. The local
spectra across the spot are then computed by the codes {\sc Titan}
and {\sc Noar} (\cite[Dumont \etal\ 2000, 2003]{dumont2000}). We
assume a black hole mass $M = 10^7 {\rm M}_\odot$, a disk accretion
rate $\dot m = 0.02$ (in units of the Eddington accretion rate), and
$r = 18 \, R_{\rm g}$ (with $R_{\rm g} = GM/c^2$).

Based on the local spectra, the relativistic ray-tracing code {\sc Ky}
(\cite[Dov{\v c}iak \etal\ 2004]{dovciak2004}) computes the time
evolution of the spectrum seen by a distant observer. We set the disk
inclination to $i = 30^\circ$ and assume a Kerr parameter $a/M =
0.998$ having known characteristics of MCG-6-30-15 in mind. In {\sc
Ky} we define the time-dependent disk emission according to the
subsequent illumination and fade-out of the spot from its center
toward the border. The duration of this time sequence is normalized by
$H$, which we constrain from the observed soft-to-hard time delay of
$\sim  600$~s (\cite[Ponti \etal\ 2004]{ponti2004}). We assume that
this delay is entirely due to the light traveling time between the
source and the disk leading to $H = 7 \, R_{\rm g}$. The flare lasts for
2000~sec, which at $r = 18 \, R_{\rm g}$ corresponds to $1/12$ of the
Keplerian time scale. More details about this type of flare model are
given in \cite[Goosmann\ (2006)]{goosmann2006}.

\begin{figure}
 \centering
 \includegraphics[width=\textwidth]{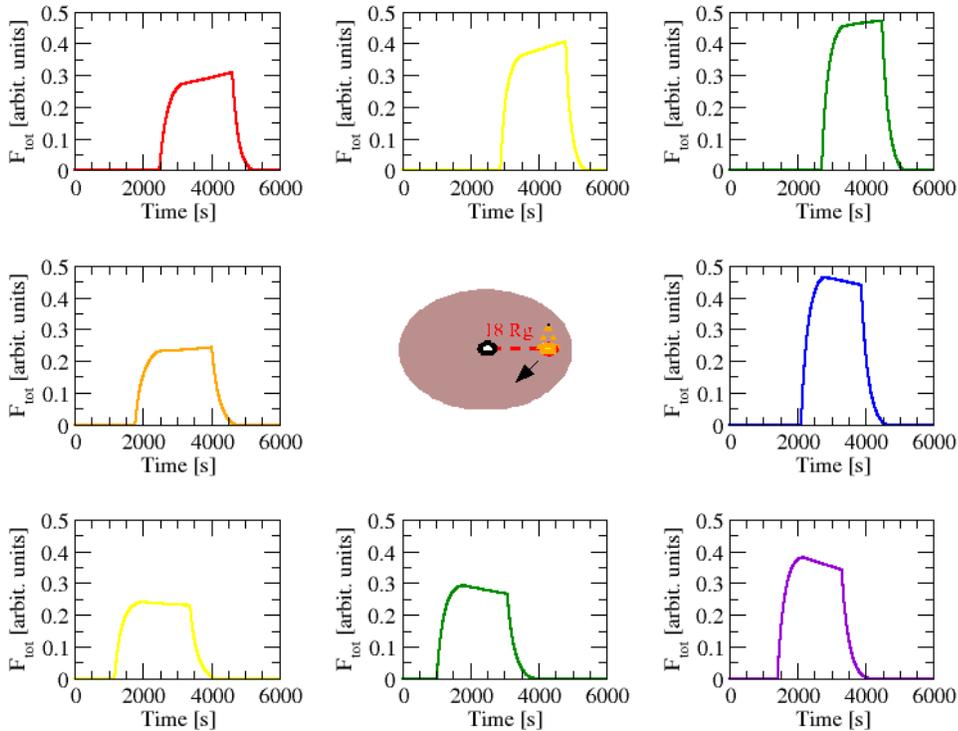}
 \caption{Lightcurves of the reprocessed spectrum integrated between 2
 keV and 10 keV for a flare lasting 2000~sec occurring at different
 orbital phases. The position of the distant observer is toward the
 bottom. \label{fig:LCs}}
\end{figure}

In figure~\ref{fig:LCs} we show the model lightcurves obtained for flares at 8
different orbital positions. They show significant differences in shape and in
normalization when comparing different orbital phases. All curves have a
broadened maximum which does not match the observed, peaked shape (see Fig.~4
in \cite[Ponti \etal\ 2004]{ponti2004}). Thus, even if the flare is
reflection-dominated, neither the geometrically evolving irradiation across
the spot nor relativistic and Doppler modifications can account for the
observed shape of the lightcurve. An intrinsic time evolution of the primary
source is therefore required.

\section{Modeling spectral time delays for a flare}

\cite[Ponti \etal\ (2004)]{ponti2004} conducted a timing analysis for
the observed flare and computed cross-correlation functions and time
delays between six different energy bands $\Delta E_{\rm i}$ (see their
Fig.~6). We attempt to reproduce the observed time delays using the
following, simple mathematical parameterization for the primary radiation
$I_{\rm{}p}$ and the reprocessed component $I_{\rm{}r}$:

\begin{eqnarray}
  I_{\rm{}p}(E,t) & = & \mathcal{L}_{\rm{}p}(t) E^{-\alpha_{\rm{}p}},\\
  I_{\rm{}r}(E,t) & = & N \mathcal{L}_{\rm{}r}(t_0,\delta,T_{\rm f},b)
  E^{-\alpha_{\rm{}r}},
\end{eqnarray}

The spectral shapes are thus represented by power laws with indexes
$\alpha_{\rm p}$ and $\alpha_{\rm r}$. The time modulations
$\mathcal{L}_{\rm{}p}(t)$ and $\mathcal{L}_{\rm{}r}(t)$ are defined by:

\begin{equation}
  \mathcal{L}_{\rm{}p}(t) = \frac{T_{\rm f}^2}{(t-t_0)^2+T_{\rm f}^2},
  \; \; \; \;
  \mathcal{L}_{\rm{}r}(t) = \frac{b^2 T_{\rm f}^2}{[t-(t_0-\delta)]^2+b^2
  T_{\rm f}^2}.
\end{equation}

These terms account for the intrinsic change of the primary that is
suggested by the modeling presented in section~\ref{sec:LCs}. The
overall factor $N$ normalizes $I_{\rm{}r}$ against the primary and
$\delta$ denotes an intrinsic delay between the two components. The
time evolution of the reprocessing can be broadened by choosing $b >
1$. This represents a spread of the light travel-time between the
source and the reprocessing site. For further details on this
model see \cite[Goosmann \etal\ (2007)]{goosmann2007}.

A distant observer detects the sum $I_{\rm{}obs}$ of $I_{\rm{}p}$ and
$I_{\rm{}r}$:

\begin{equation}
  I_{\rm{}obs}(E,t) = I_{\rm{}p}(E,t) + I_{\rm{}r}(E,t).
\end{equation}

To investigate the time delay between the spectral response of the
energy bands $\Delta E_{\rm i}$ and $\Delta E_{\rm j}$ we compute
cross-correlation functions

\begin{equation}
  F_{\rm{}CCF}^{\rm ij}(\tau) = \frac{\int
  \limits_{-\tau_{\rm{}max}}^{+\tau_{\rm{}max}} L_{\rm i}(t) L_{\rm j}
  (t-\tau) dt} {\sqrt{\int \limits_{-\tau_{\rm{}max}}^{+\tau_{\rm{}max}}
  L_{\rm i}^2(t')\;dt'} \times \sqrt{\int
  \limits_{-\tau_{\rm{}max}}^{+\tau_{\rm{}max}} L_{\rm j}^2(t'')\;dt''}},
\end{equation}

where the lightcurve $L_{\rm i}$ belongs to the energy bin $\Delta
E_{\rm i}$. We adopt here the very same method as was used in the
analysis of \cite[Ponti \etal\ (2004)]{ponti2004}. The values for
$t_0 = 500$~sec and $T_{\rm f} = 2200$~sec are adjusted to the flare
lightcurve plotted in Fig.~4 of \cite[Ponti \etal\ (2004)]{ponti2004}.

%\begin{table}\def~{\hphantom{0}}
%  \begin{center}
%  \caption{Energy bands and fixed parameters of the delay model}
%  \label{tab:kd}
%  \begin{tabular}{lccc}
%      \hline
%      Energy band & Range [keV] & Fixed parameter & Value\\
%      \hline
%      $\Delta E_1$ & 0.2 -- 0.57 & $E_{\rm{}min}$ & 0.2 keV\\
%      $\Delta E_2$ & 0.6 -- 2.2  & $E_{\rm{}max}$ & 12 keV\\
%      $\Delta E_3$ & 2.2 -- 2.6  & $\alpha_{\rm{}p}$ & 1.3\\
%      $\Delta E_4$ & 2.6 -- 4.8  & $t_0$ & 2200 {\rm s}\\
%      $\Delta E_5$ & 4.8 -- 6.8  & $T_{\rm f}$   & 500 s\\
%      $\Delta E_6$ & 6.8 -- 12.0 & ~ &~ \\
%      \hline
%  \end{tabular}
% \end{center}
% \label{tab:param}
%\end{table}

The spectral slope of the reprocessed component has a major impact on the
obtained energy-dependent time delays. We choose a rather
hard slope of $I_{\rm{}r}$ and set $\alpha_{\rm r} = 0.1$. Such a
spectral shape corresponds to the reprocessing in a medium of magnetically
confined, cold clouds as analyzed by \cite[Kuncic, Celotti, \& Rees\
(1997)]{kuncic1997}. Their Fig.~2 shows reprocessed model spectra
that are much harder than the primary due to multiple soft X-ray absorption by
the individual clouds. For the primary slope we set $\alpha_{\rm p} =
1.3$ having in mind that for the higher flux state during the flare
the spectrum should steepen (\cite[Shih, Iwasawa, \& Fabian\
 2002]{shih2002}). 

\begin{figure}
  \centering
  \includegraphics[width=0.58\textwidth]{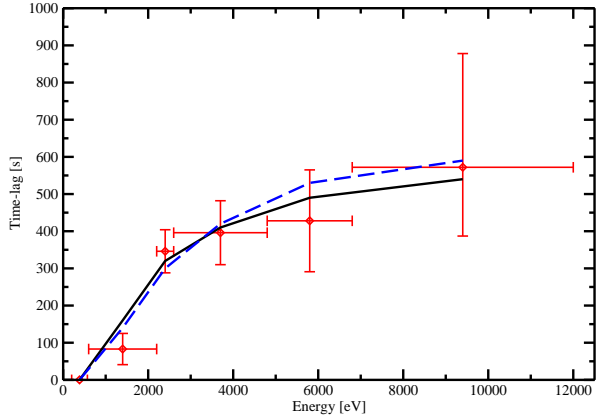}
  \caption{Best representation of the measured time-lags (red diamonds with
  error bars) for the flare/clumps scenario. The three curves rely on the
  following parameterization: $\delta = 1000$~sec, $b = 4$ and $N = 0.9$ (solid
  black line) and $\delta = 900$~sec, $b = 3$ and $N = 0.6$ (dashed blue
  line). \label{fig:mcg-delay-fit}}
\end{figure}

The model can reproduce the observed time lags for the observed flare of
MCG-6-30-15. In Fig.~\ref{fig:mcg-delay-fit} we show two examples of
satisfactory data representation. The results account for the slight
flattening of the delay curve toward higher energies. It turns out that
different choices of parameters can lead to similar delay curves, as
there are three free parameters involved ($N$, $b$, and $\delta$) and
the error bars of the data are large. To distinguish between different
parameter sets some fine tuning of the delay curve`s shape would be
necessary. This is not meaningful with the current data.

The observed time-delay cannot be reproduced when the spectral slope
of $I_{\rm{}r}$ is softer, i.e. adjusted to an ionized reflection
scenario as proposed by \cite[Ballantyne, Vaughan, \& Fabian\
(2003)]{ballantyne2003}. If the primary source is located within $\sim
10 \, R_{\rm g}$ of the black hole, the relative variability of the
primary and the reflection component can be reproduced (\cite[Miniutti
\& Fabian\ 2004]{miniutti2004}). Such a setup also explains the high
equivalent width of the iron K$\alpha$ line, which requires a
relatively large amount of reflection. While these ionized reflection and
light-bending models are certainly important to explain the behavior of
MCG-6-30-15 on longer time-scales, the data obtained during the 2000~sec flare
period does not allow to conclude on the time evolution of the iron K$\alpha$
line. From the results of our delay modeling, we therefore suggest the
possibility that the strong individual flare does not originate in the inner
accretion flow, where most of the X-ray energy is dissipated. We rather
imagine that the flare occurred farther away from the black hole, in a region
where the disk is fragmented and replaced by magnetically confined, cold
clumps. 

\acknowledgements We are grateful to A.-M. Dumont and A. R{\'o}{\.z}a{\'n}ska
for their help with computing the local spectra used in this model.

\end{document}